# Magnetic Properties of epitaxial Re/Co$_{1-x}$Au$_x$/Pt heterostructures

Sukanta Kumar Jena[1,2,3,4,*], Anuj Kumar Dhiman[5,6,7], Artem Lynnyk [1], Kilian Lenz[2], Gauravkumar Ishwarbhai Patel[2], Aleksiej Pietruczik[1], Paweł Aleszkiewicz[1], Jürgen Lindner[2], Piotr Dłużewski[1], Ryszard Gieniusz[5], Andrzej Maziewski[5], Ewelina Milińska[1], Andrzej Wawro[1]

1. Institute of Physics, Polish Academy of Sciences, Warsaw, Poland
2. Institute of Ion Beam Physics and Materials Research, Helmholtz-Zentrum Dresden-Rossendorf, Germany
3. Jerzy Haber Institute of Catalysis and Surface Chemistry, Polish Academy of Sciences, Krakow, Poland
4. Bilakankuda, Kapilewsarpur, Puri, Odisha, Bharat
5. Faculty of Physics, University of Bialystok, Bialystok, Poland
6. Faculty of Physics, Adam Mickiewicz University, Poznan, Poland
7. Max Plank Institute for Polymer Research, Mainz, Germany

*sukaphysics@gmail.com

**Abstract:**

We investigate epitaxial Co(20 Å) and Co$_{1-x}$Au$_x$(20 Å) alloy thin-films surrounded by asymmetric heavy metals layers of Re(10 Å) as a buffer and Pt(30 Å) as a cap to study the magnetic anisotropy, interfacial Dzyaloshinskii-Moriya interaction (iDMI) and damping. The increase of Au from 0% to 25% in the Co$_{1-x}$Au$_x$ alloy generates the spin-reorientation transition of around 13% of Au. The increase in Au concentration provides a significant decrease in saturation magnetization from 1690 kA/m to 982 kA/m measured for Co and Co$_{75}$Au$_{25}$, respectively. The effective anisotropy constant $K_{eff}$ is elevated up to 0.33 MJ/m$^3$ by changing the Au content. Further, our investigations of the magnetization dynamics have confirmed that the overall effective damping constant rises with the Au concentration which can be attributed to the spin pumping effect. The spin pumping leads to the highest value of effective spin mixing conductance $g^{\uparrow\downarrow} \approx 2.91 \times 10^{18}/m^2$ in the Co$_{90}$Au$_{10}$(20 Å) system, while the lowest value of $g^{\uparrow\downarrow} \approx 2.25 \times 10^{18}/m^2$ is found for the Co(20 Å) system. Additionally, we have investigated the iDMI strength, and the amplitude of iDMI decreases with increasing Au concentration. The highest surface iDMI constant value equal to 2.62 pJ/m is observed for Co.

**Introduction:**

Over the past two decades, researchers have focused on identifying the most efficient materials that can be sandwiched with heavy metals to exhibit perpendicular magnetic anisotropy (PMA), chirality in domains, and high damping constant for spintronic applications. These exceptional materials offer the possibility for fast switching in magnetic spintronic devices, which plays an important role in the advancement of high-density magnetic data storage technologies with low power consumption[1,2]. The PMA originates from the magneto-crystalline anisotropy in a specific crystallographic direction depending upon the crystal field and the spin-orbit interaction (SOI) [13]. The advantage of the hexagonal-closed packed structure of Co is that it offers PMA along the *c*-axis direction (called the easy axis). In typical magnetic thin films as used in spin-orbitronic devices, such as random access memory devices based on spin transfer torque[4], the PMA of Co is influenced by the adjacent buffer and capping layers. The interfaces between the ferromagnet and the heavy metals play a very important role in inducing the PMA in multilayer heterostructures. Furthermore, the PMA is also influenced by the crystalline quality and interfacial roughness of the sample. In graphene/Co/Pt trilayers, it was demonstrated that the Co exhibits the highest level of PMA with the spin re-orientation transition (SRT) up to 40 Å[5]. It is well known that the growth of Co in the surrounding of an

Au or Pt layer (buffer or cap) can vary between hcp or *fcc* structure, and depending on this structure, a certain thickness of PMA can be achieved [6,7]. Usually, the SRT occurs at 19 Å for hcp Co neighboring with Pt layers (Pt/Co/Pt)[8]. However, when the capping layer is substituted with Au, W, or Mo, the SRT is shifted towards lower thicknesses depending on the heavy metals used in the capping layer[7,9]. Recently, it has been shown that W for the buffer layer in Pt/W($t_W$)/Co/Pt multilayers plays a very important role in enhancing the SRT thickness of Co, and the SRT thickness increases with increasing W thickness ($t_W$)[9]. Additionally, a rapid reduction in PMA was found when Co becomes thicker at constant W thickness, which leads to the ultimate loss of its interfacial properties. It has been found that the $Co_{1-x}Au_x$ alloy is typically immiscible in bulk phases[10]. Moreover, the magnetic layer thickness for PMA can be enhanced when the Au concentration increases up to a certain extent in the $Co_{1-x}Au_x$ alloy[11]. In this context, emphasizing on the higher PMA, we study the magnetic behavior of $Al_2O_3$(0001)/Pt (400 Å)/Re (10 Å)/$Co_{1-x}Au_x$ (20 Å)/Pt (30 Å) alloys, where the Au concentration is varied from $x = 0$ to 25%. The few available literature studies confirm either the amorphous nature of the $Co_{1-x}Au_x$ alloy or the existence of two phases, namely Co and $Co_{1-x}Au_x$ nano alloys[11,12]. However, we discuss the crystallinity of bimetallic $Co_{1-x}Au_x$ alloy using reflection high energy electron diffraction (RHEED) patterns and cross-sectional transmission electron microscopy (TEM). Additionally, we have investigated the dynamic behavior of Re(10 Å)/$Co_{1-x}Au_x$(20 Å)/Pt(30 Å) heterostructures to determine the effective damping, spin pumping, and effective spin mixing conductance from ferromagnetic resonance (FMR) spectroscopy.

Recently, it has been reported that the Au atoms gain an additional magnetic moment when Au is added to the Co system[13] similar to the interfacial Au in Co/Au multilayers[14], which is responsible for the enhanced magnetic damping. It is well-known that materials with lower damping are better candidates for memory and spintronic applications as well as for electromagnetic shielding. The lower damping materials are used for the purpose of lower current density in spintronics devices[15,16]. In contrast, the materials with higher damping can be used in magnetic memory devices for fast writing and reading the data as it reduces the time of magnetization switching[17,18,19,20,21].

By engineering the neighboring layers of Co and $Co_{1-x}Au_x$ alloys with heavy metals (HM) as buffer and cap, it is possible to induce the interfacial Dzyaloshinskii-Moriya interaction (iDMI) at the interface between the ferromagnetic (FM) and the HM, which is very useful for forming chiral structures like magnetic skyrmions [22,23,24]. To stabilize the skyrmions or chiral domains, it is important to optimize both, the PMA and the iDMI amplitude. The estimation method for the iDMI amplitude is crucial, especially in the case of magnetic layers containing few monolayers. One of the reliable and straightforward methods to estimate the iDMI involves the asymmetric spin wave propagation, which can be measured by Brillouin light scattering (BLS) spectroscopy. It is sensitive in detecting the asymmetric spin-wave propagation in the magnetic thin films with asymmetric buffer and capping layers[25]. We determined the iDMI by including the saturation magnetization taken from superconducting quantum interference device (SQUID).

Recently, it has been shown that sputtered Pt/Co/Re[26] gives the highest surface iDMI constant $D_S$ as compared to other sputtered metallic multilayers, such as Pt/Co/Ir [26,27,28] W/Co/Pt[29], and Pt/Co/Ta [25]. However, our molecular beam epitaxy (MBE)-grown samples with pure Co exhibit higher surface iDMI ($D_S$) values compared to the sputtered multilayer systems [26]. Therefore, the tuning of both, higher damping and iDMI amplitude by Au concentration, makes AuCo alloy a promising candidate for magnetic storage devices.

**Experimental Details and Methods:**

We fabricated the samples by MBE with a deposition pressure of $10^{-8}$ mbar. First, we deposited a 400-Å-thick Pt buffer on an $Al_2O_3(0001)$ substrate at a growth temperature of 850 °C. Then the temperature was reduced to room temperature to deposit the Re(10 Å)/$Co_{1-x}Au_x$(20 Å)/Pt(30 Å) trilayers. The $Co_{1-x}Au_x$ (20 Å) films were prepared by co-deposition of Co and Au from effusion cells with the desired concentration. The thickness was controlled by a quartz crystal balance. During the growth, we monitored the RHEED patterns to observe the epitaxial nature of the $Co_{1-x}Au_x$ alloy. The Re and $Co_{1-x}Au_x$(20Å) alloy grow pseudomorphically, i.e., both follow the structure of the Pt buffer. The Pt cap layer protects the sample from oxidation and plays an important role in inducing the iDMI. All samples are listed in the Table 1.

Table I: List of the fabricated samples.

| Sample name | Sample details |
| --- | --- |
| S1 | $Al_2O_3(0001)/Pt(400Å)/Re(10Å)/$**$Co(20Å)$**$/Pt(30Å)$ |
| S2 | $Al_2O_3(0001)/Pt(400Å)/Re(10Å)/$**$Co_{95}Au_5(20Å)$**$/Pt(30Å)$ |
| S3 | $Al_2O_3(0001)/Pt(400Å)/Re(10Å)/$**$Co_{90}Au_{10}(20Å)$**$/Pt(30Å)$ |
| S4 | $Al_2O_3(0001)/Pt(400Å)/Re(10Å)/$**$Co_{85}Au_{15}(20Å)$**$/Pt(30Å)$ |
| S5 | $Al_2O_3(0001)/Pt(400Å)/Re(10Å)/$**$Co_{80}Au_{20}(20Å)$**$/Pt(30Å)$ |
| S6 | $Al_2O_3(0001)/Pt(400Å)/Re(10Å)/$**$Co_{75}Au_{25}(20Å)$**$/Pt(30Å)$ |
| S7 | $Al_2O_3(0001)/Pt(400Å)/$**$Re(10,20,30Å)$**$/$**$Co_{95}Au_5(0-30Å)$**$/Pt(30Å)$ |
| S8 | $Al_2O_3(0001)/Pt(400Å)/$**$Re(10,20,30Å)$**$/$**$Co_{90}Au_{10}(0-30Å)$**$/Pt(30Å)$ |
| S9 | $Al_2O_3(0001)/Pt(400Å)/$**$Re(10,20,30Å)$**$/$**$Co_{80}Au_{20}(0-30Å)$**$/Pt(30Å)$ |

The Co/heavy metals interfaces Re/Co and Co/Pt exhibit similar iDMI signs [26,27]. Hence, we assume that the same applies to the Re/$Co_{1-x}Au_x$ and $Co_{1-x}Au_x$/Pt interfaces for $x>0$ as well.

For the structural analysis, we performed cross-sectional TEM. We measured the magnetic parameters by SQUID magnetometry. Further, we measured the dynamics by vector-network-analyzer FMR in a frequency range of 5 to 45 GHz with steps of 1 GHz using a coplanar waveguide (CPW) with an 80-μm-wide centre conductor. The magnetic field was perpendicular to the sample surface during measurement, i.e., we performed out-of-plane measurements for all six samples S1-S6, where the Au concentration varies from 0 to 25%. We measured the FMR in field-sweep mode at constant frequency. The FMR spectra were fitted by a complex Lorentzian to determine the peak-to-peak linewidth and the resonance field. We employed BLS in backscattering geometry where the samples were measured in Damon-Eshbach configuration i.e., the external magnetic field oriented perpendicular to the wave vector ($k$). By illuminating the sample with a monochromatic continuous wave green laser ($\lambda$ = 532 nm), data is collected using a tandem Fabry-Perot interferometer. We calculated the iDMI constant from the $k$-dependence of the frequency difference between Stokes and anti-Stokes peaks.

Note that the static and dynamic measurements were performed for samples S1 to S6 (as given in Table I). However, the magneto-optical Kerr-effect magnetometry (MOKE) measurements at remanence were performed for samples S7 to S9.

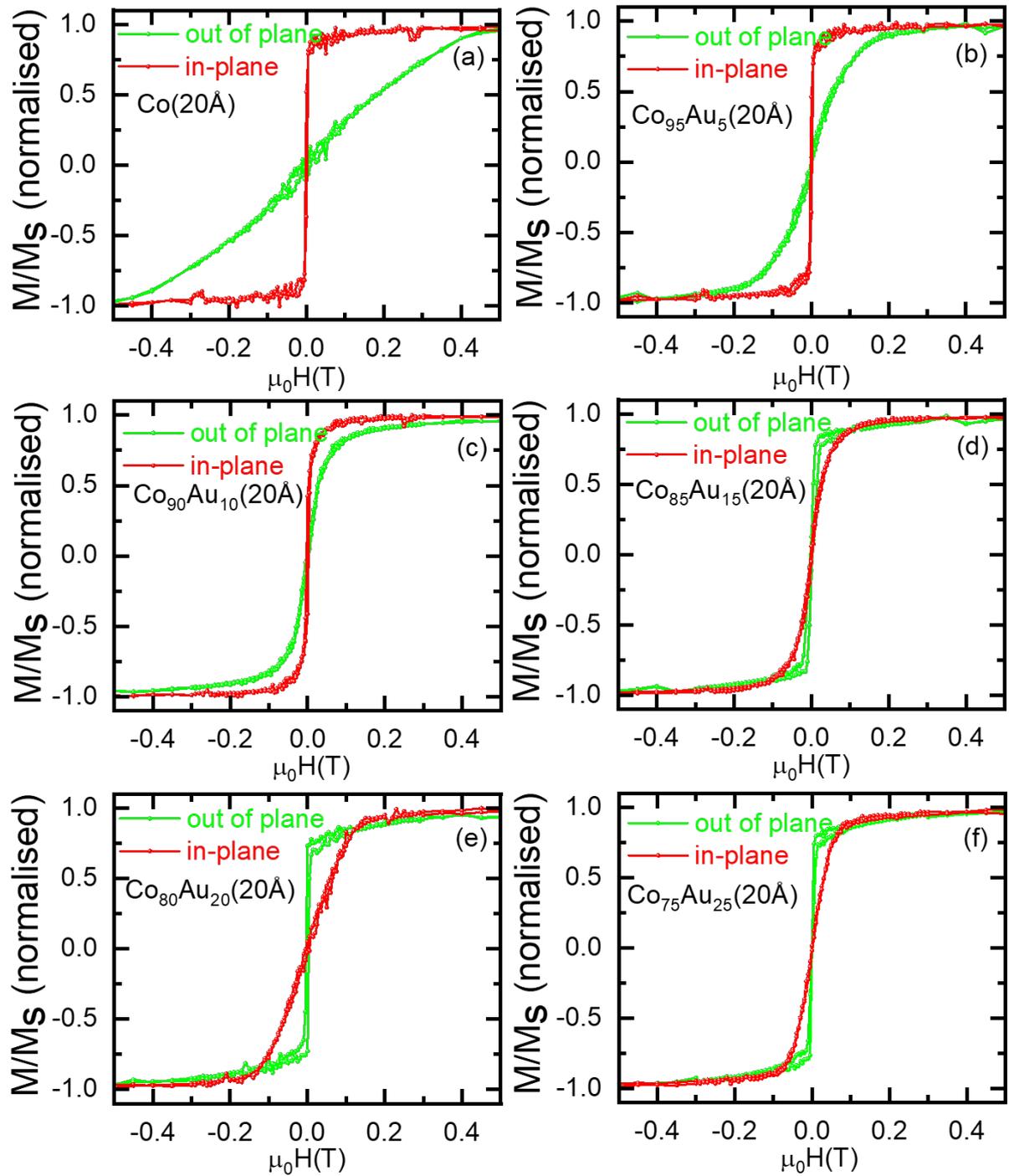

*Figure 1: Magnetization reversal loops measured by SQUID for out-of-plane (green) and in-plane (red) orientation of the 20-Å-thick samples S1–S6 with Au concentrations from 0% to 25%. The out-of-plane anisotropy dominates the in-plane anisotropy from 10% to higher Au content.*

**Results and Discussion:**

From the SQUID measurements of samples S1-S6 (see ***Fig. 1***) we determined the saturation magnetization $M_S$. It decreases with increasing Au concentrations (see Table I). The in-plane reversal loop for the sample with Au≤10% shows a square hysteresis with non-zero but very low coercivity. The square reversal loops turn into S-shaped loops with zero coercivity as the Au concentration increases from 10% to 15%. We observe the opposite behavior for the out-of-plane loops of samples with Au concentrations raising from 10% to 15%. It is important to note that both in-plane and out-of-plane loops should have zero coercivity at the transition point from in-plane anisotropy to out-of-plane easy axis (see **Fig. 1**).

The effective anisotropy ($K_{eff}$) is determined from the enclosed area between the in-plane and out-of-plane hysteresis loop. Afterward, the uniaxial anisotropy constant ($K_U$) is calculated from ($K_{eff}$) using the saturation magnetization by $K_U = \frac{\mu_0 M_S^2}{2} + K_{eff}$. **Fig. 2(a)** exhibits the $K_U$ (red spheres) as a function of Au concentration. $K_U$ decreases as the concentration of Co decreases i.e., Au concentration increases. Further increase in Au content increases $K_{eff}$ as shown in **Fig. 2(a)**. The result is listed in table II.

Table-II: Magnetic parameters retrieved from FMR, SQUID, and BLS data. We extracted the parameters $\mu_0 M_{eff}, g, \mu_0 \Delta H$ and $\alpha$ from the FMR experimental data of $H_{res}$ vs. $f$ and $\Delta H_{pp}$ vs. $f$ by using equations (1) and (2), and $D_{eff}$ and $D_S$ from BLS experimental data using equation (4), respectively.

| Sample | VNA-FMR parameters | | | | | SQUID parameters | | | BLS parameters | |
|---|---|---|---|---|---|---|---|---|---|---|
| | $\mu_0 M_{eff}$ (T) | $K_U$ (kJ/m³) | $g$ | $\alpha$ | $\mu_0 \Delta H$ (mT) | $M_S$ (kA/m) | $K_{eff}$ (kJ/m³) | $K_U$ (kJ/m³) | $D_{eff}$ (mJ/m²) | $D_S$ (pJ/m) |
| S1 | -0.43769 | 1425 | 2.13 | 0.0242 | 25.4 | 1690 | -295.7 | 1499 | 1.31 | 2.62 |
| S2 | -0.17134 | 1104 | 2.21 | 0.0304 | 20.6 | 1395 | -61.4 | 1162 | 0.99 | 1.98 |
| S3 | -0.05312 | 1085 | 2.20 | 0.0333 | 17.1 | 1335 | -74.2 | 1047 | 1.12 | 2.24 |
| S4 | 0.03153 | 898 | 2.21 | 0.0345 | 25.6 | 1182 | 20.5 | 899 | 0.93 | 1.87 |
| S5 | 0.11121 | 808 | 2.19 | 0.0369 | 6.7 | 1090 | 25.5 | 773 | 0.76 | 1.52 |
| S6 | 0.06191 | 637 | 2.20 | 0.0386 | 13.4 | 982 | 38.4 | 645 | 0.57 | 1.14 |

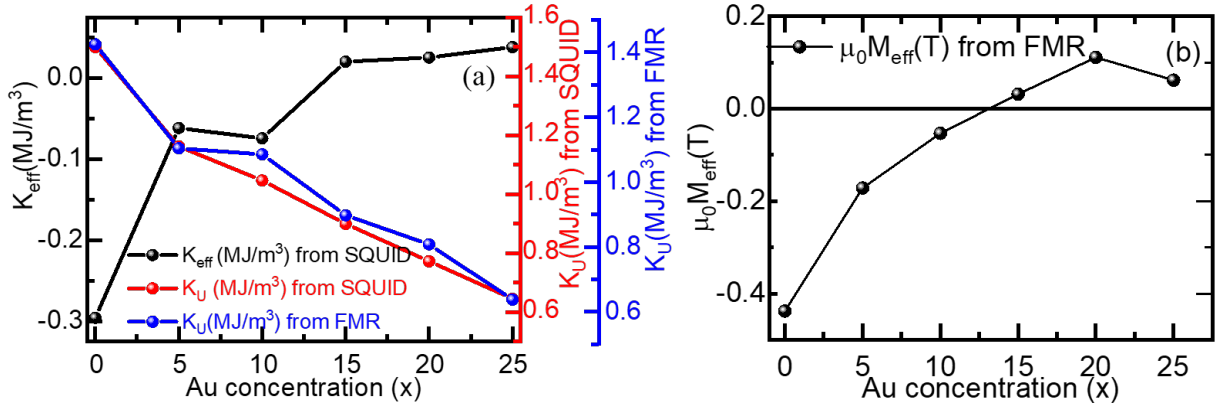

*Figure 2: Au-concentration dependence of (a) $K_{eff}$ (left) determined from MH loop and Ku (right) determined from SQUID (red) and FMR (blue), and (b) effective magnetization from FMR.*

We measured the room temperature FMR to evaluate the *g*-factor, effective magnetization, and effective damping. **Fig. SM1** in the supplementary materials section shows the fitted frequency-field dependence. The solid line depicts the fit according to Kittel's resonance condition for a cubic system in a perpendicular direction. The resonance condition with $\theta_H = 0°$ reads[30]:

$$f = \frac{\gamma}{2\pi}(\mu_0 H_{res} - \mu_0 M_{eff}) \qquad (1)$$

where, $f$ is the microwave frequency, $\gamma = \frac{g\mu_B}{\hbar}$ is the gyromagnetic ratio with the Landé $g$ factor, $\mu_B$ is the Bohr magneton, $\mu_0 H_{res}$ is the resonance field and $\mu_0 M_{eff}$ is the effective magnetization. Further, we fitted the frequency-dependence of the linewidth, to yield the total effective Gilbert damping and the inhomogeneous broadening term $\mu_0 \Delta H_0$. The equation for Gilbert damping is given by[31]:

$$\mu_0 \Delta H_{pp} = \mu_0 \Delta H_0 + \frac{2\alpha}{\sqrt{3}\gamma}\omega \qquad (2)$$

where $\omega = 2\pi f$ and $\mu_0 \Delta H_{pp}$ is the peak-to-peak linewidth. The peak-to-peak linewidth vs. the applied frequency is shown in the supplementary materials *Fig. SM2* for the samples S1-S6.

We observed that the inhomogeneous broadening of the linewidth ($\Delta H_0$) decreases with increasing the Au concentration for samples S1, S2, and S3 [see Table II and *Fig. 5(b)*], which suggests that the interface is getting better with increasing the doping of Au. However, a deviation occurs for S4 sample, where the inhomogeneous broadening again increases and it further decreases for samples S5 and S6. *Fig. 2(b)* shows that the effective magnetization $\mu_0 M_{eff}$ decreases for samples S1-S6. The effective magnetization exhibits the lowest value of -0.17 T for 5% of Au doping, then increases to 0.11 T for 25% Au. This observation indicates that the transition from easy plane to easy axis occurs as the sign of effective magnetization changes from positive to negative at around 13% Au.

Further, we study the details of the effective damping of samples S1 to S6. The damping is a very crucial parameter for the dynamic properties of magnetic metals. The damping stems not only from the magnetic material but also other effects such as interfacial roughness[32], extrinsic damping, spin pumping, and magnetic proximity effect (MPE) which occurs due to the presence of neighbouring HM layers (e.g., Re and Pt at Co interface in our case), play an important role in damping enhancement. Moreover, the presence of Pt at the interface causes spin pumping[33], and if thick enough, increases the effective damping. In our $Co_{1-x}Au_x$ alloy samples, the damping increases with the Au content as mentioned in the *Fig. 5(a)*. This phenomenon can be explained by the following reasons.

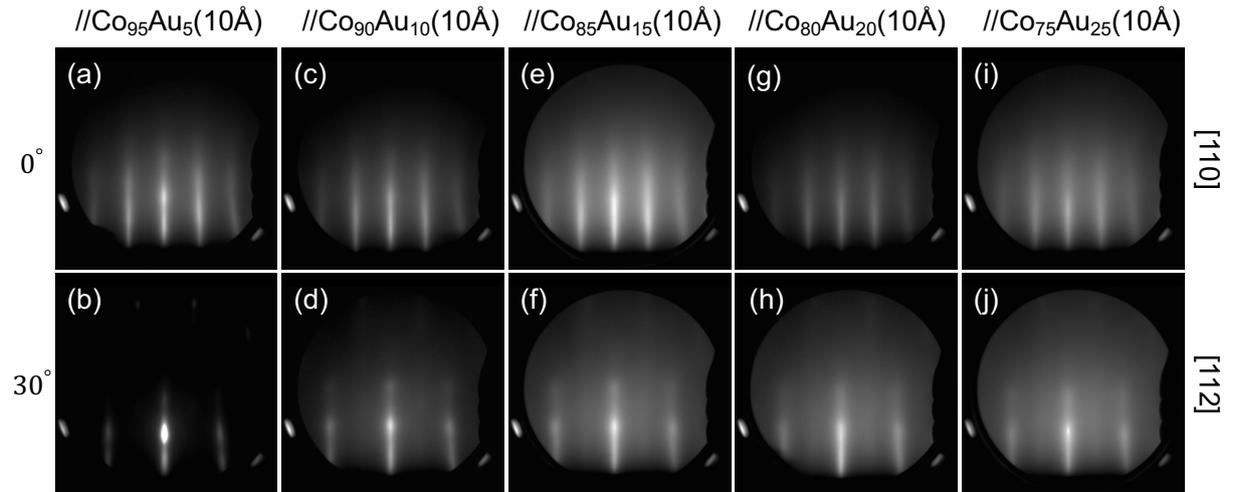

Figure 3: RHEED images taken during the growth of the $Co_{1-x}Au_x$ alloy films. The upper row and lower row images show the (110) and (112) planes, respectively.

One of the factors is that with the increase of the Au concentration, the disorder in Co might increase. Subsequently, this increases the damping. We performed an in-situ RHEED analysis during the sample growth to investigate the disorder in crystal quality i.e., to check if there is any agglomeration of Co or Au nanocrystal in $Co_{1-x}Au_x$ alloy, and to deduce the epitaxial relation of $Co_{1-x}Au_x$ with respect to the Re layers. ***Fig. 3*** shows the RHEED patterns for the different Au contents for 0° and 30° corresponding to the (110) and (112) planes, respectively. The RHEED streaks for samples containing Au concentrations in the range of 15% to 25% are a bit blurry and less distinct than those observed for less Au and in pure Co (see ***Fig. 3***). It suggests that the inclusion of Au in the $Co_{1-x}Au_x$ alloys might have a detrimental effect on the crystalline quality. However, the RHEED measurement confirms that all $Co_{1-x}Au_x$ samples possess a six-fold rotational geometrical symmetry.

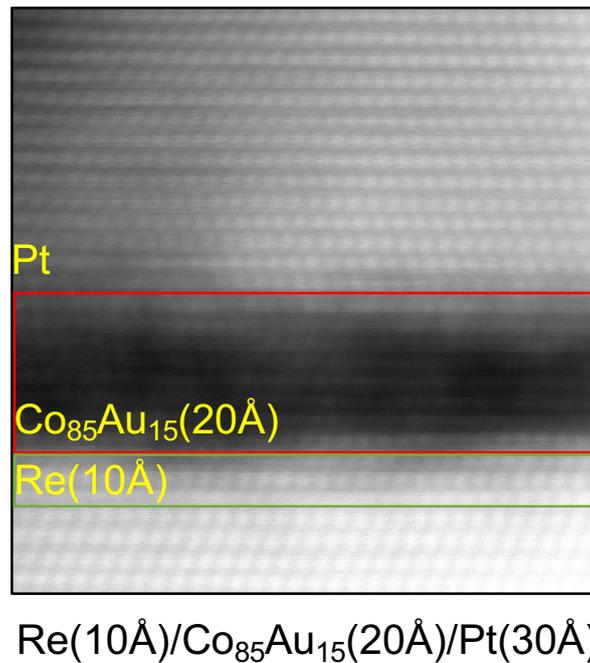

Re(10Å)/Co$_{85}$Au$_{15}$(20Å)/Pt(30Å)

*Figure 4: Cross-sectional TEM image for sample S4 ($Co_{85}Au_{15}$) with the various layers indicated.*

To check the interfacial roughness and the crystal quality, cross-sectional TEM imaging was performed for sample S4. ***Fig. 4*** shows a sharp interface to Re and the growth is homogeneous, unlike the island type growth in sputtered samples. The $Co_{85}Au_{15}$ is crystalline with the atomic layers aligned parallel to the Re buffer lattice suggesting the epitaxial growth of Pt, Re, and $Co_{85}Au_{15}$ alloy. As discussed above, it can be expected that the Au-doping may have a negative effect on the crystal structure due to the higher structural disorder. However, the interface is sharp, as confirmed by TEM, which suggests that the interfacial roughness is negligible. From this perspective, the structural disorder or interfacial roughness is not a relevant factor to cause larger effective damping when the Au concentration is increased. Therefore, the larger effective damping in the $Co_{1-x}Au_x$ alloys should depend on other factors such as MPE or spin pumping, which we will discuss below.

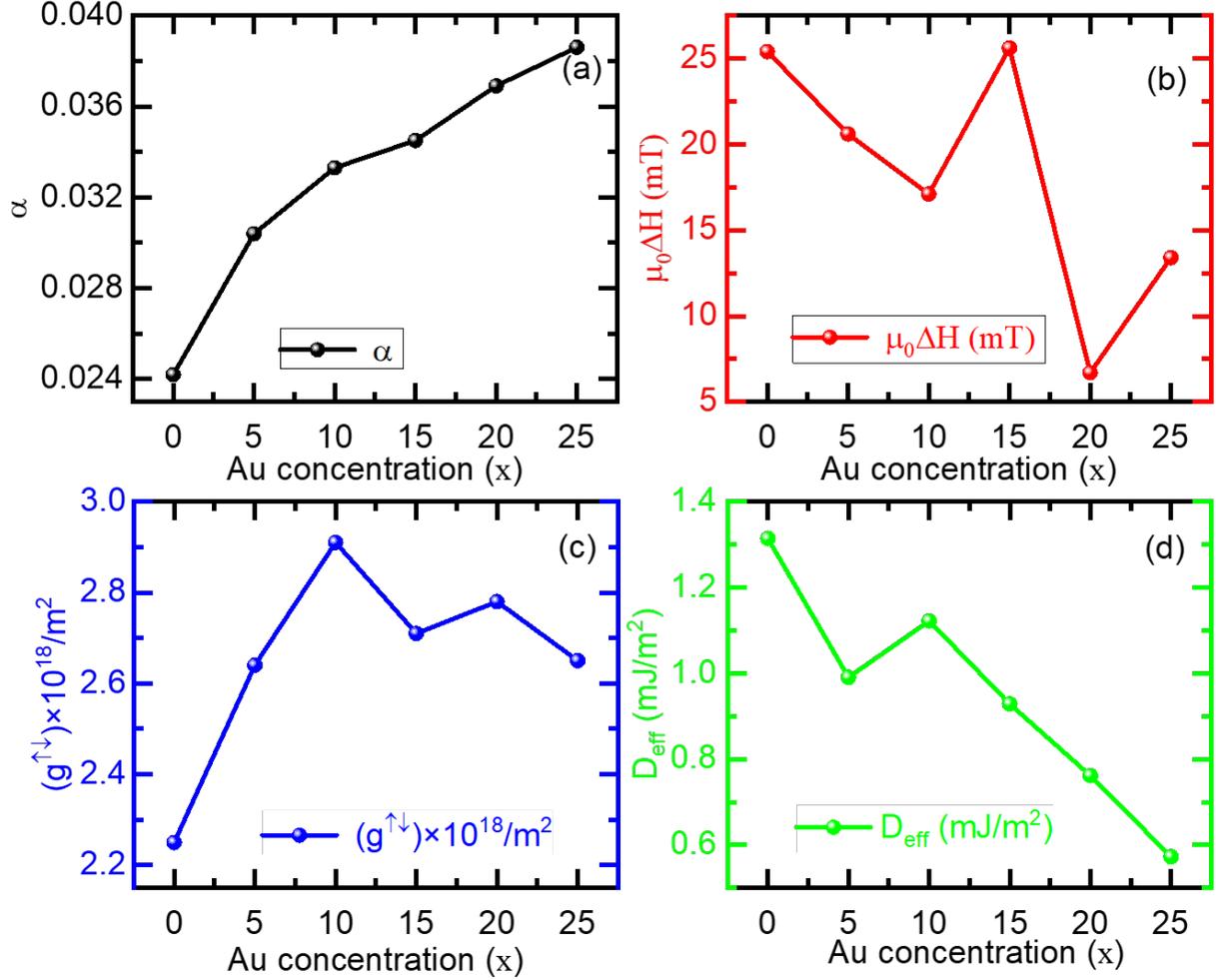

*Figure 5: Au concentration dependence of (a) effective damping (α) (b) inhomogeneous broadening of linewidth (μ$_0$ΔH$_0$), (c) spin mixing conductance (g$^{\uparrow\downarrow}$), and (d) effective DMI D$_{eff}$.*

One potential reason for the higher damping might be the proximity-induced magnetization of Au atoms by their neighboring Co atoms[13]. With increasing Au content, more proximity-induced magnetization occurs, which can directly influence the effective damping constant of the Co$_{1-x}$Au$_x$ alloys through spin pumping. The effective damping is a combination of intrinsic and extrinsic contributions, whereby the intrinsic components consist of Gilbert damping and interfacial damping. The main factor of intrinsic damping is the electron scattering due to phonons and magnons[34,35]. With respect to the electron scattering, the magneto-crystalline anisotropy and SOI are the important parameters ruling the intrinsic damping[36,37]. The interfacial damping depends on the interface quality and the thickness of the magnetic materials. Here, it is well known that with decreasing thickness to transit from in-plane to out-of-plane magnetic anisotropy, the effective damping constant increases. Nevertheless, in the samples we studied here, we consider the interfacial damping as negligible owing to the presence of well-defined interfaces. The extrinsic damping consists of various contributions among which the two-magnon scattering effect is a prominent factor[38]. However, for the out-of-plane FMR measurements in our systems, the two-magnon scattering is not active. Other contributions to the extrinsic damping are the spin pumping effect, arising from the dissipation of the FMR-generated spin current in the Pt or Re layers, and the MPE due to *d-d* hybridization inducing a ferromagnetic order in the Pt up to a few atomic layers at the interface[39,40]. However, it is difficult to quantify the damping due to MPE. In the Rojas-Sanchez *et al.* [39], Cu was employed as a spacer in between Co and Pt to eliminate or minimize the MPE. This observation revealed a reduction of spin pumping, damping, and spin Hall angle. Similarly, Cocan *et al.* utilized MgO as a spacer layer between Fe and Pt and compared results with Fe/Al[41]. Interestingly, they observed that the damping of Fe/Al is higher than Fe/MgO/Pt, which was attributed to the variations of the interface and different growth behavior arising from the buffer and cap layer. However, it is crucial not to disregard but to consider the MPE effect contribution to the effective damping. Apart from the aforementioned factors, there are further contributions, such as eddy current damping and radiative damping, which add additional complexity to the effective damping[42,43].

The FMR measurements reveal a significant 25.61% increase in the damping for Au concentrations of 0 to 5% as shown in ***Fig. 5(a)***. However, for 10% to 25% Au, we observe only a gradual and modest rise of 4-9 % in the effective damping. Next, we determine the spin-mixing constant ($g^{\uparrow\downarrow}$) to quantify the spin pumping effect with respect to the total damping. The amplitude of $g^{\uparrow\downarrow}$ can be estimated by[44]:

$$\alpha - \alpha_0 = \frac{g\mu_B}{\mu_0 M_S t_{FM}} g^{\uparrow\downarrow} \tag{3}$$

where, $\alpha$ is the effective damping determined by FMR, $\alpha_0$ is the damping constant for the bulk value of Co equal to 0.011[45], and $t_{FM}$ is the thickness of the ferromagnetic material, i.e. the $Co_{1-x}Au_x$ layer. The effective spin-mixing conductance describes the total spin current transfer at the FM/HM interface, which is dissipated in the HM or partially reflected. The values of $g^{\uparrow\downarrow}$ are given in Table III, where the lowest value of $g^{\uparrow\downarrow} \approx 2.25 \times 10^{18}/m^2$ is observed for the pure Co sample S1.

Table III: Effective spin mixing conductance ($g^{\uparrow\downarrow}$) extracted for S1-S6 by using equation (3).

| Sample | $g^{\uparrow\downarrow} \times 10^{18}/m^2$ |
|---|---|
| S1 | 2.25 |
| S2 | 2.64 |
| S3 | 2.91 |
| S4 | 2.71 |
| S5 | 2.78 |
| S6 | 2.65 |

Doping Au into Co increases the effective $g^{\uparrow\downarrow}$ and reaches the highest value for sample S3 with $g^{\uparrow\downarrow} \approx 2.91 \times 10^{18}/m^2$ as shown in ***Fig. 5(c)***. Hence, the additional spin pumping causes the increase of the effective damping. For higher Au concentrations (S4, S5, and S6) $g^{\uparrow\downarrow}$ decreases again. This observation leads to the conclusion that an enhancement of the effective damping cannot be due to spin pumping only. Contributing factors such as damping due to MPE, eddy current, and radiative damping might play a significant role here. Concurrently, the effective $g^{\uparrow\downarrow}$ value increases while the inhomogeneous linewidth broadening decreases (see ***Fig. 5(b) & (c)***. It is a fine interplay between the improved interface, which on the one hand reduces inhomogeneity but on the other hand increases $g^{\uparrow\downarrow}$. As a result, the increase in spin pumping at these improved interfaces further increases the effective damping constant of the system. Therefore, the inclusion of Au in Co promotes a smoother interface and promotes a stronger PMA. Notably, the transition from in-plane easy axis to out-of-plane easy axis may also contribute to the larger effective damping.

Further, we investigated the samples by BLS to explore the DMI at the interfaces between Re/Co$_{1-x}$Au$_x$ and Co$_{1-x}$Au$_x$/Pt. The asymmetrical interaction arises due to the presence of high SOC and lack of inversion symmetry. The presence of the aforementioned two factors allows for quantifying the accurate strength of the iDMI. We measure the Stokes ($f_S$) and anti-Stokes ($f_{aS}$) peaks related to the spin-wave modes propagating in opposite directions of the wave vector $k$. From the frequency asymmetry, $\Delta f=|f_S-f_{aS}|$, for Damon-Eshbach modes, it is possible to extract the effective DMI constant by:

$$D_{\text{eff}} = \Delta f \frac{\pi M_S}{2\gamma\, k} \tag{4}$$

We list the values in Table II. The corresponding surface (interfacial) iDMI constant $D_S$ can be defined as $D_{eff} \times t_{FM}$.

The iDMI constant for pure Co (sample S1), $D_S$ = 2.62 pJ/m, is slightly higher than that for the epitaxial Pt/Co/Cu multilayer system with $D_S$ is equal to 2.4 pJ/m determined from micromagnetic simulations[46]. However, for the epitaxial trilayer Cu/Co/Pd systems, the value of $D_S$ is equal to 0.11pJ/m determined from domain wall motion[47]. Similarly, the epitaxial Pt/Co/Au trilayer exhibits the lower $D_S$ of 0.6pJ/m [48]. Jena et. al. reported the strength of $D_S$ equal to 1.83pJ/m for epitaxial W/Co/Pt multilayers [29] estimated from the K$_{eff}$ method. For sputtered systems:- Ir/Co/Pt ($D_S$ = 0.96 pJ/m)[28], and Pt/Co/Re ($D_S$ = 2.1 pJ/m)[26], we found that the strength of $D_S$ is lower than epitaxial Re/Co/Pt system. The above discussion shows that the Ds depends upon the interface and the crystal structure of heavy metals with high spin orbit coupling strength. It should be also noted that the higher iDMI is due to the additive nature of both interface contributions, namely from the Re/Co and the Co/Pt interfaces[26]. The above discussed Ds for various systems are lower than the Ds value of 2.62pJ/m for Re/Co/Pt stack determined by BLS. To our knowledge, there are no literature describing iDMI study of Co$_{1-x}$Au$_x$ alloy with different asymmetric interfaces.

With increasing Au concentration, the iDMI decreases in the Co$_{1-x}$Au$_x$ alloy as shown in the *Fig. 5(d)*. Please note that the iDMI is directly proportional to the *Δf*. Here, we observed the linear correlation between iDMI amplitude and saturation magnetisation value. As a result, the iDMI amplitude decreses with decreasing the Ms value at the same thickness but different Au concentrations of Co$_{1-x}$Au$_x$ alloy. Here, in our case of epitaxial samples and sharp interfaces, the strength of $D_{eff}$ and $D_S$ are higher compared to the sputtered amorphous and multi-crystalline trilayers found in literatures[26,28,29].

In the sections above, we discussed the static and dynamic properties of our samples having a particular thickness of the magnetic layers. Now, we want to discuss another aspect of the Co$_{1-x}$Au$_x$ alloy, namely the dependence of the SRT on the sample thickness and Re buffer thickness. As mentioned earlier, the magnetic thickness at which the transition from the out-of-plane easy axis to the in-plane easy axis is referred to as the SRT. In our Co$_{1-x}$Au$_x$ alloy system, the increasing Au concentration leads to a larger SRT thickness. In spintronic devices, a higher thickness with an out-of-plane easy axis is desired. So, tuning the Au concentration in Co$_{1-x}$Au$_x$, the SRT as well as the saturation magnetization can be tailored. *Fig. 6* illustrates the variation of the SRT for different Au concentrations (5%, 10% and 20%). If the Au concentration remains fixed, increasing the Re buffer layer thickness results in a proportional change in the SRT thickness. This phenomenon can be attributed to the following mechanism.

In the case of a pure Co thin film, the SRT thickness is constant for Re(10 Å)/Co($t_{FM}$)/Pt(30 Å) trilayers. By adding Au to in the $Co_{1-x}Au_x$ alloy, the Co concentration decreases. Therefore, all magnetic moments of Co in $Co_{1-x}Au_x$ alloy continue to align in the perpendicular direction at the SRT thickness. The higher thickness of $Co_{1-x}Au_x$ alloy is necessary to exhibit the SRT, which is still PMA in higher thickness compared to pure Co thin films. When the thickness of the $Co_{1-x}Au_x$ alloy exceeds the SRT of pure Co, the atomic moments of Co strive to align the whole magnetic moment from out-of-plane to in-plane direction resulting in the occurrence of the SRT.

For thicker Re buffers, i.e. 20 and 30 Å compared to 10 Å, the $Co_{1-x}Au_x$ alloy shows a higher SRT thickness. This may be explained in terms of interfacial roughness[49,50]. It is well-known that elevated growth temperatures[50] are required to grow epitaxial Re thin films having a minimum surface roughness. In contrast, when Re is deposited at room temperature, the surface roughness may be comparatively higher [49]. We found that for 10Å of Re bottom layer, the interface between Re and $Co_xAu_{x-1}$ alloy is very sharp as confirmed by TEM. There is possible increase in the roughness for the higher Re layer thickness because of room temperature deposition. Therefore, the interfacial roughness may affect the various magnetic properties to some extent.

In our samples S1 to S9, we grew Re at room temperature, which gives a hcp structure [45], [46] and has a six-fold rotational geometrical symmetry as confirmed by RHEED (see supplementary **Fig. SM4**). However, from the RHEED images, it is difficult to distinguish the *fcc* and *hcp* symmetry. Again, the lattice constant is the same as for Pt confirmed from RHEED streaks (see supplementary materials **Fig. SM3 (a and b)**). It can be expected that the formation of dead layers could be induced at the Re/FM interface due to the presence of interfacial roughness of Re, which converts magnetic layers into nonmagnetic (dead) layers for higher thicknesses of Re. It is also possible that with increasing the Re thickness, the grain size increases[50], which may be responsible for creating the magnetic dead layer of $Co_{1-x}Au_x$ at the Re/ $Co_{1-x}Au_x$ interface. The remaining magnetic layer of $Co_{1-x}Au_x$ alloy (excluding the dead layer) is not sufficient to cause an SRT. As a result, it affects the magnetic SRT. Therefore, the higher thickness of the magnetic layer is necessary to exhibit the SRT for specific concentrations of Au.

From the RHEED analysis [as shown in the supplementary materials **Fig. SM3(e), (f), (g), (h), and (i)**], it is confirmed that the lattice constant of the $Co_{1-x}Au_x$ alloy increases with rising Au concentration. As a result, the higher lattice constant of $Co_{1-x}Au_x$ alloy gives a higher SRT thickness as Co exhibits a higher lattice constant with Au than for Mo [7].

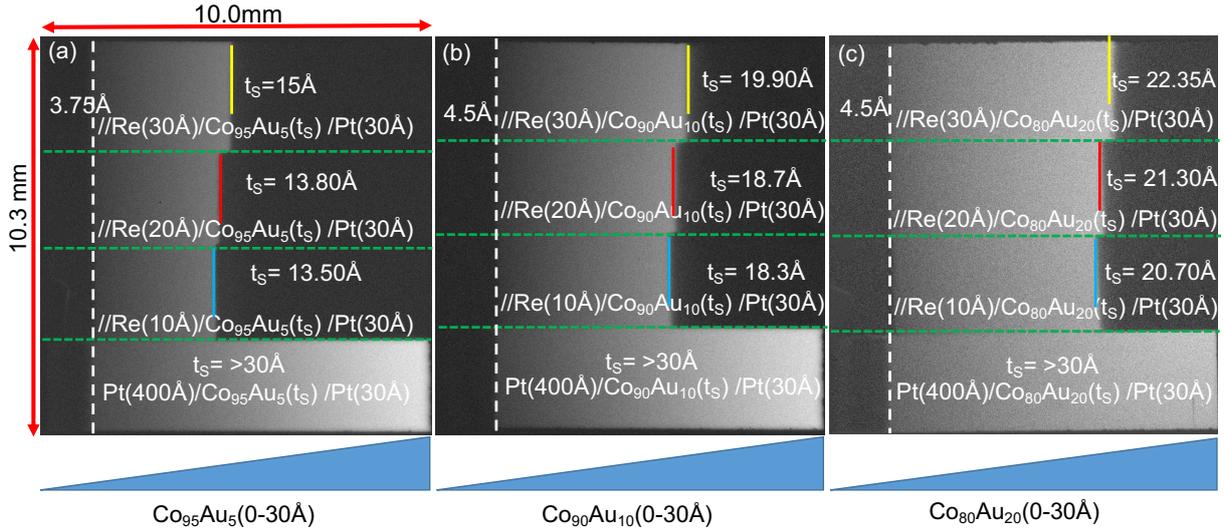

*Figure 6: Kerr microscopy images taken in remanence showing the SRT thicknesses (dark/bright contrast change) of (a) $Co_{95}Au_5$, (b) $Co_{90}Au_{10}$, and (c) $Co_{80}Au_{20}$ films for Re buffer thicknesses of 30 Å, 20 Å, 10 Å, and 0 Å corresponding to a 400 Å Pt buffer, respectively. The dashed white lines in the left mark the transition from a superparamagnetic state to the ferromagnetic state. The horizontal green dashed lines represent the different Re buffer thickness zones.*

*Fig. 6(a)* illustrates four different zones, which consist of different thicknesses of the Re buffer layer (0, 10, 20, and 30 Å) for a particular Au concentration: $Co_{95}Au_5$. Without a Re buffer, i.e. just directly on the 400-Å-thick Pt layer, the thickness for $Co_{95}Au_5$ SRT is equal to 31.5Å. This means that the SRT thickness is larger than Co SRT (as shown in the supplementary figure SM6). Using a 10-Å-thick Re buffer between Pt and $Co_{1-x}Au_x$, the SRT thickness decreases sharply down to 13.5 Å. The dark contrast on the right side in each image of *Fig. 6* represents the in-plane easy axis orientation, whereas the bright contrast represents the out-of-plane easy axis orientation. Moreover, the SRT thickness for the $Co_{95}Au_5$ alloy increases with increasing Re buffer thickness, as depicted in *Fig. 6(a)*. This trend is observed similarly for $Co_{90}Au_{10}$ (*Fig. 6(b)*) and $Co_{80}Au_{20}$ (*Fig. 6(c)*), where the magnetic layer thickness needed for the SRT increases even further. When there is no Re buffer layer, but only Pt cap and buffer layers, the SRT is increased with Au concentration and it is 41.3 Å for $Co_{80}Au_{20}$ which is slightly higher than the graphene/Co/Pt system[5]. In summary, when a specific value of Re thickness is considered, the SRT thickness shows an upward trend with increasing Au concentration. In other words, the SRT thickness of the magnetic layers increases with increasing Au concentration in the $Co_{1-x}Au_x$ alloys.

**Summary**:

We investigated the various magnetic properties of $Co_{1-x}Au_x$ alloys. We revealed that with increasing Au concentration in the $Co_{1-x}Au_x$ alloy, the saturation magnetization decreases. The saturation magnetization for 5% Au is still nearly equal to the standard Co bulk value, which suggests that additional magnetic moment might be induced due to addition of Au and proximity effect at Pt interface. However, there is a sharp reduction of $M_s$ by the inclusion of 5% Au. FMR measurement confirmed that the effective damping constant increases with Au content. We attribute this mainly to the spin-pumping phenomena. The iDMI amplitude increases with decreasing Au concentration. It leads to the highest value of 1.31 mJ/m$^2$ for Re(10 Å)/Co(20 Å)/Pt(20 Å), while Re(10 Å)/$Co_{75}Au_{25}$(20 Å)/Pt(20 Å) shows the lowest value of 0.57 mJ/m$^2$. As the Au concentration in $Co_{1-x}Au_x$ alloy increases, the SRT thickness increases for a fixed Re buffer thickness and capping layer. The SRT shifts towards higher thicknesses when the Re buffer layer thickness is increased given a fixed Au concentration and $Co_{1-x}Au_x$ thickness. From the RHEED and TEM analysis, we revealed that the $Co_{1-x}Au_x$ exhibits inherent crystalline properties and its growth is epitaxial following the structure of the Re buffer. Our study gives a new path to achieve the stabilization of $Co_{1-x}Au_x$ bimetallic alloys at room temperature through epitaxial growth using MBE, which is typically difficult due to the inherent immiscibility of this alloy in bulk phase. Therefore, it could be useful for not only fundamental research but also for spintronics applications in the future. By adjusting the Au concentration, it is possible to tailor the different important parameters such as PMA, damping, and iDMI.

**Acknowledgement:**

SKJ and AKD acknowledge student seed funding by the IEEE magnetics society 2020. This work was supported by the Foundation for Polish Science (FNP) under the European Regional Development Fund – Program [REINTEGRATION 2017 OPIE 14-20] and by the Polish National Science Centre projects: [2016/23/G/ST3/04196] and OPUS-19 [2020/37/B/ST5/02299]. AW would like to acknowledge the M-ERA.NET 3 (2022/04/Y/ST5/00164). The authors acknowledge Braj Bhusan Singh from Harcourt Butler Technical University, Bharat, for his assistance in discussing SOT.

**Author contributions:**
SKJ, AW, and EM proposed the research project. SKJ leads the project. SKJ and AL have performed the SQUID and its analysis. SKJ and AP deposited the samples and analysed the RHEED images. SKJ, GKP, JL and KL performed the VNA-FMR and analysis. PA and AP performed the Kerr microscopy. RG, AKD, AM performed BLS and its analysis. PD performed the cross-sectional TEM and its analysis. SKJ and AKD wrote the manuscript. All authors contributed and revised the manuscript and discussed the results.

**Conflict of interest**
The authors declare that they have no conflict of interest.